\documentclass[a4paper]{appolb}
\usepackage{graphicx}
\usepackage{amsmath}
\usepackage{multirow}
\usepackage{cite}

\newcommand{\xscrpt}[1]{{\textstyle \mbox{\scriptsize #1}}}

\begin{document}
\title{Problems with meson spectroscopy involving perturbative loop corrections}
\author{K. P. Khemchandani
\address{Instituto de F\'isica, Universidade de S\~ao Paulo, C.P. 66318, 05389-970 S\~ao 
Paulo, SP, Brazil}\\\vspace{0.3cm}
Eef Van Beveren
\address{$^{2}$Centro de F\'{\i}sica Computacional,
Departamento de F\'{\i}sica, Universidade de Coimbra, P-3004-516 Coimbra, Portugal}\\\vspace{0.3cm}
George Rupp
\address{$^{3}$Centro de F\'{\i}sica das Interac\c{c}\~{o}es Fundamentais,
Instituto Superior T\'{e}cnico, Universidade de Lisboa, 
P-1049-001 Lisboa, Portugal}}
\maketitle
\begin{abstract}
In this talk we review the limitations of including meson loops as perturbative
corrections in a solvable quark model. We first discuss meson-meson scattering
within a
formalism which treats confined quark pairs and mesons on an equal footing. The
interaction between the mesons proceeds through $s$-channel meson-exchange
diagrams.  Next, we develop a perturbative expansion of the model, and show
that the resonance poles found in such a treatment, even by accounting for
contributions up to fourth order, do not coincide with those obtained with the
full model. We conclude that the resonance predictions based on perturbative
approximations in quark models are not reliable, especially in those cases
where the coupling to the scattering channels is large.
\end{abstract}

\section{Introduction}
One of the main challenges currently faced in the field of hadron physics is
to understand the hadron spectra. Observation of more and more new hadronic
states is being reported continuously, many of which do not seem to fit into
the traditional quark spectrum of $q\bar{q}$ mesons and $qqq$ baryons. In fact,
evidence of the existence of some hadrons has recently been found in
experimental data whose quantum numbers essentially indicate nontraditional
quark configurations. For example, several hadrons, labelled as the
``$Z_c$/$Z_b$'' states, with charmonium/bottomonium-like mass but nonzero
electric charge, seem to have been discovered \cite{reviewz}. Such states
necessarily require more than two constituent quarks to get their quantum
numbers right. Curiously, many of these new hadrons lie very close to the
threshold  of some open charm/bottom meson systems, which suggests
that it might be possible to understand them as bound states or Feshbach
resonances in systems of open-charm/bottom mesons. In fact, this possibility
has been successfully explored within various models \cite{hosaka,marina,us}.

Actually, indications of an exotic nature for various hadrons has been under
discussion since several decades, with one of the first cases possibly being
that of the light scalar mesons \cite{ZPC30p615}, whose structure is still
subject to debate.

The difficulty with studies related to hadron spectroscopy arises due to the
fact that perturbative QCD cannot be used  at low energies and one has to
depend on models to interpret the experimental data. It is important to
remember that experimental evidence of possible resonances is obtained from
total or partial-wave cross sections, as well as angular distributions and
decay modes. Thus, theoretical analysis plays a crucial role in extracting
information from the data, and models have to be built wisely and carefully. 
In some occasions it is even possible that an alternative explanation of
experimental data exists, such as threshold effects and cusps
\cite{evb,swanson}, or phase space with different angular momentum
\cite{alberto}.

On other occasions it is important to see if certain assumptions in the model
can lead to unreliable results. In the present manuscript, we focus on the
errors resulting from considering meson loops in perturbative approximations
in quark models.

\section{Formalism}
We first discuss our model for studying meson-meson scattering, which is based
on treating quarks and hadrons as coupled systems within a nonperturbative
formalism.  The basic idea behind the model is to consider the
string-breaking feature of the confining force. Accordingly, at a certain
separation the ``string'' between a quark and an antiquark in a meson will be
energetically favoured to break, followed by the creation of a new and light
$q\bar{q}$ pair, which then may lead to hadronic decay. This possibility is
considered in the present model by writing the decaying meson in terms of a
confined $q\bar{q}$ pair, which along with a new quark pair generated by the
$3P_0$ mechanism rearranges to form a system of two mesons. This process,
together with the time-reversed one, leads to meson exchange in the
$s$-channel. Thus, we consider the
exchange of a quark pair with radial quantum number varying from 0 to infinity,
as shown in Fig.~\ref{fig1}:
\begin{figure}[h!]
\centering
\includegraphics[width=0.6\textwidth]{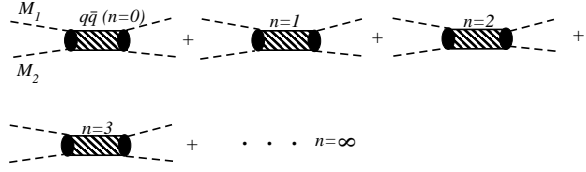}
\caption{The meson-meson potential,
which involves the $s$-channel exchange of a confined $q\bar{q}$ pair, 
with radial quantum number $n$
running from 0 to $\infty$.}\label{fig1}
\end{figure}

Mathematically, the model is written
in terms of a coupled system of nonrelativistic Hamiltonians
\begin{eqnarray}
H_{c}\psi_{c}(\vec{r})+V_{T}(\vec{r})\psi_{f}(\vec{r}) &=&
E\psi_{c}(\vec{r}),
\label{hc}\\
H_{f}\psi_{f}(\vec{r})+V_{T}(\vec{r})\psi_{c}(\vec{r}) &=&
E\psi_{f}(\vec{r}),
\label{hf}
\end{eqnarray}
where the subscripts ``$c$'' and ``$f$'' (here and throughout this article)
refer to the confined quarks and free mesons
(i.e., considering them plane waves), respectively, and $V_{T}$ is the
transition potential between the two sectors. Furthermore,
$H_{c}$ and $H_{f}$ stand for the Hamiltonians of these sectors. The confining
potential is assumed to be a harmonic oscillator, viz.\ 
\begin{equation}
V_{c}=\frac{1}{2}\mu_{c}\omega^{2}r^{2},
\end{equation}
with $\mu_{c}$ and $\omega$ the reduced mass and frequency
of the $q\bar{q}$ system, respectively. 
We denote the energy eigenvalue of $H_{c}$ by $E_{nl}$,
i.e.,
\begin{equation}
E_{nl}=\omega (n_{c}+l_{c}+3/2) + m_{q} + m_{\bar{q}}.
\label{enl}
\end{equation}
For the transition potential, we take a local delta-shell function of the form
\begin{equation}
\langle\vec{r}_{f}\mid V_{T}\mid\vec{r}_{f}^{\prime}\rangle
= \frac{\lambda}{\mu_{c}a}\delta (r_{f}- a)
\delta^{3}(\vec{r}_{f}-\vec{r}_{f}^{\prime}).
\label{vt}
\end{equation}
This form of potential has been proven useful
in describing the breaking of the colour string
\cite{IJTPGTNO11p179}.
The $\lambda$ and $a$ in Eq.~(\ref{vt}) are the two parameters of the model,
with the former being the coupling of the meson-meson channel
to the $q\bar{q}$ channel, and the latter the average interquark distance
for string breaking.
The coupling $\lambda$ is varied between 0 and 1 in the present study,
with $\lambda=0$ corresponding to decoupled two-meson and quark-antiquark
systems. Since the meson-meson state is considered a plane wave,
decoupling would result in a pure (``bare'') $q\bar{q}$ spectrum.
On the other hand, $\lambda\,\geq$ 1 represents
strong coupling to the meson-meson channel.
The parameter $a$ is taken roughly in the range
1.7--2.5~GeV$^{-1}$ (0.34--0.5 fm).

Using these ingredients it is possible to obtain  a simple closed-form
expression for the $S$-matrix, in the case that only one confined and one free
channel are considered (see Appendices A.1--A.5
of Ref.~\cite{IJTPGTNO11p179} for a detailed derivation), viz.\
\begin{equation}
S_{l_{f}}(E) = 1 - 2i\frac{2a\lambda^{2}\sum\limits_{n=0}^\infty
\dfrac{g_{n}^{2}}{E(\vec{P_{f}})-E_{nl}}\mu_{f}P_{f}j_{l_{f}}
(P_{f}a) h_{l_{f}}^{1}(P_{f}a)}
{1+2ia\lambda^{2}\sum\limits_{n=0}^\infty
\dfrac{g_{n}^{2}}{E(\vec{P_{f}})-E_{nl}}
\mu_{f}P_{f}j_{l_{f}}(P_{f}a) h_{l_{f}}^{1}(P_{f}a)}.
\label{smat}
\end{equation}
An exact solution for the $S$-matrix can be derived in the most general 
multichannel case as well, resulting in a matrix expression with a 
similar structure \cite{PRD80p094011}.

In order to find resonances in non-exotic meson-meson systems,
we search for zeros in the denominator of the $S$-matrix (Eq.~(\ref{smat})),
which correspond to poles in the complex-energy plane. Then we vary the
parameters $\lambda$ and $a$ to fit the experimental data for a particular
case. 

The main purpose of this manuscript is to show that predictions
of resonance poles based on perturbative calculations in standard quark
models can be very misleading. In order to demonstrate this in a quantitative
way, we construct a perturbative scheme to include the meson loops.

As explained above, $\lambda$ in our formalism
is the coupling of the meson-meson $\leftrightarrow\, q\bar{q}$ vertex.
The term corresponding to $\lambda^{2}$ thus represents
the lowest-order meson-meson interaction
(meson-meson $\rightarrow\, q\bar{q} \, \rightarrow\,$ meson-meson).
The position of a pole in a particular case of meson-meson scattering,
above threshold, can be expanded perturbatively in terms of $\lambda^{2}$ as
\begin{equation}
E_{m}^\xscrpt{pole}= E_{m} +\lambda^{2} E_{m}^\xscrpt{LO}
+\lambda^{4} E_{m}^\xscrpt{NLO}
+\lambda^{6} E_{m}^\xscrpt{NNLO}+\,\ldots .
\label{empole}
\end{equation}
The first term of this series corresponds to the confinement pole,
which is what we should get from the model if the coupling of the quark pair
to the meson channels vanished.
The second term is the leading-order term in $\lambda^{2}$, which corresponds
to including a one-meson-loop correction.
The denominator of the $S$-matrix (Eq.~(\ref{smat})), written
up to leading order in $\lambda^{2}$
around the $m$th confinement pole, becomes
\begin{equation}
1+2ia\lambda^{2}\frac{g_{m}^{2}}
{E_{m}+\lambda^{2} E_{m}^\xscrpt{LO}-E_{m}}
\mu k j_l(ka) h_l^{1}(ka) = 0,
\end{equation}
which then yields
\begin{equation}
E_{m}^\xscrpt{LO}= - 2ia g_{m}^{2}\mu k j_l(ka) h_l^{1}(ka).
\end{equation}
So the pole position (Eq.~(\ref{empole})),
in lowest-order approximation, is then
\begin{equation}
E_{m}^\xscrpt{pole}\approx E_{m}-2ia g_{m}^{2}\mu k j_l(ka) h_l^{1}(ka).
\end{equation}
Similarly, one can obtain the expressions for even higher-order contributions
(see Ref.~\cite{ptp_us} for more details).

To test the validity of the perturbative calculus, we now choose an example of
a meson-meson system, namely $K\pi$ in a $P$ wave, where we
study the $K^*(892)$ pole. The $K^{\ast}(892)$ vector meson is well
described by a Breit-Wigner resonance in $P$-wave $K\pi$ scattering,
with central mass and resonance width of about 892 MeV
and 50 MeV, respectively.
Hence, we expect a pole in the $S$-matrix of Eq.~(\ref{smat})
for an $S$-wave nonstrange-strange quark-antiquark system,
coupled to a $P$-wave kaon-pion meson-meson system.

\begin{figure}[htbp]
\centering
\includegraphics[scale=0.45]{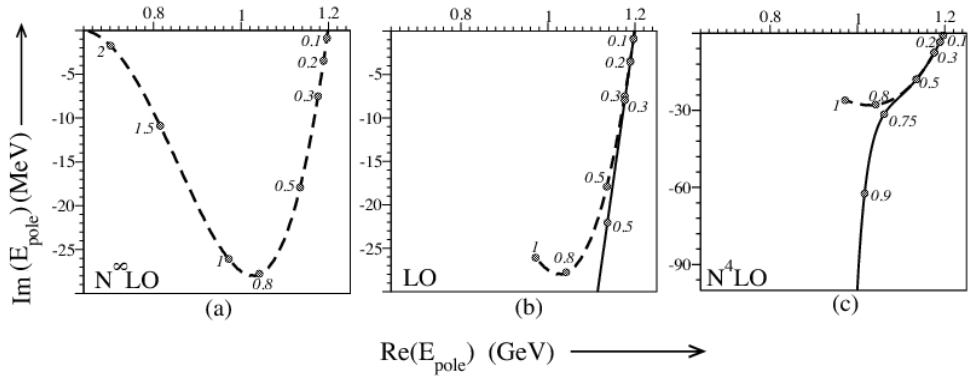}\vspace{-3.5cm}
\caption{\small
The $P$-wave $K\pi$ resonance pole positions as a function of
the coupling $\lambda$.
The dashed curves for N$^{n}$LO ($n=0, 4$)
are the same as the one shown for N$^{\infty}$LO (labelled (a)).
The solid curves are the results obtained
from the  perturbative approximations, viz.\
(b) leading (N$^{0}$LO, Born) term,
and (c) (next-to)$^{4}$-leading (N$^{4}$LO) orders, respectively.}
\label{figKpiP}
\end{figure}

Scattering poles are obtained by studying the zeros
of the denominator in the expression of Eq.~(\ref{smat}).
In Fig.~\ref{figKpiP}(a), we depict
the $S$-matrix pole positions 
for a range of  $\lambda$ values
varying from 0 to just over 2. The values of all input parameters are given in
Ref.~\cite{ptp_us} (as determined earlier in Ref.~\cite{PRD27p1527}).
 
In the limit of vanishing coupling,
one expects to find the poles at the bare masses
of the quark-antiquark system, as given by Eq.~(\ref{enl}).

We obtain from Eq.~(\ref{enl}) the value $E_{00}=1.199$ GeV for the
ground-state bare mass, which indeed corresponds to the limit of
vanishing $\lambda$ along the dashed curve in Fig.~\ref{figKpiP}(a).
The value $\lambda=1$ corresponds to the physical pole,
as it roughly reproduces the characteristics of the $K^*(892)$
resonance. In the present simplified model, the pole comes out
at $(0.972-i0.026)$~GeV, as shown in Fig.~\ref{figKpiP}(a).

In Figs.~\ref{figKpiP}(b,c) we depict
the perturbative pole trajectories for the bare nonstrange-strange
$q\bar{q}$ state at 1.199~GeV. Shown are the curves
for the lowest-order (Born) term
($E_{0}^{\xscrpt{N}^{0}\xscrpt{LO}}$)
and for the fourth-order term
($E_{0}^{\xscrpt{N}^{4}\xscrpt{LO}}$), respectively.
We find that the Born term gives satisfactory 
pole positions for overall couplings up to $\lambda\approx0.3$.
The perturbative pole position
of the $K^{\ast}(892)$ resonance is well determined 
up to $\lambda\approx0.75$ for the fourth-order approximation.
However, there above things go terribly wrong, and the approximation
completely fails to reproduce the physical pole at $\lambda=1$.

We can thus conclude that,
although the results of the perturbative expansion seem to improve somewhat
by including terms of higher and higher order, the characteristics of the
$K^{\ast}(892)$ resonance do not get reproduced at all.
Moreover, we should add that these higher-order perturbative calculations
are much more tedious than just finding the exact solution for the coupled
quark-antiquark and meson-meson system in our solvable model.

\section{Acknowledgements}
K. P. Khemchandani thanks the Brazilian funding agencies FAPESP and CNPq for
financial support.

\end{document}